\PassOptionsToPackage{unicode}{hyperref}
\PassOptionsToPackage{hyphens}{url}
\documentclass[
  11pt,
]{article}
\usepackage{lmodern}
\usepackage{amssymb,amsmath}
\usepackage{ifxetex,ifluatex}
\ifnum 0\ifxetex 1\fi\ifluatex 1\fi=0 
  \usepackage[T1]{fontenc}
  \usepackage[utf8]{inputenc}
  \usepackage{textcomp} 
\else 
  \usepackage{unicode-math}
  \defaultfontfeatures{Scale=MatchLowercase}
  \defaultfontfeatures[\rmfamily]{Ligatures=TeX,Scale=1}
\fi
\IfFileExists{upquote.sty}{\usepackage{upquote}}{}
\IfFileExists{microtype.sty}{
  \usepackage[]{microtype}
  \UseMicrotypeSet[protrusion]{basicmath} 
}{}
\makeatletter
\@ifundefined{KOMAClassName}{
  \IfFileExists{parskip.sty}{%
    \usepackage{parskip}
  }{
    \setlength{\parindent}{0pt}
    \setlength{\parskip}{6pt plus 2pt minus 1pt}}
}{
  \KOMAoptions{parskip=half}}
\makeatother
\usepackage{xcolor}
\IfFileExists{xurl.sty}{\usepackage{xurl}}{} 
\IfFileExists{bookmark.sty}{\usepackage{bookmark}}{\usepackage{hyperref}}
\hypersetup{
  pdftitle={Simulating time-to-event data from parametric distributions, custom distributions, competing risk models and general multi-state models},
  pdfauthor={Karolinska Institutet},
  hidelinks,
  pdfcreator={LaTeX via pandoc}}
\usepackage[margin=1in]{geometry}
\usepackage{color}
\usepackage{fancyvrb}

\DefineVerbatimEnvironment{Highlighting}{Verbatim}{commandchars=\\\{\}}
\usepackage{framed}
\definecolor{shadecolor}{RGB}{248,248,248}
\newenvironment{Shaded}{\begin{snugshade}}{\end{snugshade}}

\newcommand{\CommentTok}[1]{\textcolor[rgb]{0.56,0.35,0.01}{\textit{#1}}}

\newcommand{\DataTypeTok}[1]{\textcolor[rgb]{0.13,0.29,0.53}{#1}}
\newcommand{\DecValTok}[1]{\textcolor[rgb]{0.00,0.00,0.81}{#1}}

\newcommand{\FunctionTok}[1]{\textcolor[rgb]{0.00,0.00,0.00}{#1}}

\newcommand{\KeywordTok}[1]{\textcolor[rgb]{0.13,0.29,0.53}{\textbf{#1}}}
\newcommand{\NormalTok}[1]{#1}

\newcommand{\OtherTok}[1]{\textcolor[rgb]{0.56,0.35,0.01}{#1}}

\newcommand{\StringTok}[1]{\textcolor[rgb]{0.31,0.60,0.02}{#1}}

\usepackage{graphicx,grffile}
\makeatletter
\def\maxwidth{\ifdim\Gin@nat@width>\linewidth\linewidth\else\Gin@nat@width\fi}
\def\maxheight{\ifdim\Gin@nat@height>\textheight\textheight\else\Gin@nat@height\fi}
\makeatother
\setkeys{Gin}{width=\maxwidth,height=\maxheight,keepaspectratio}
\makeatletter
\def\fps@figure{htbp}
\makeatother
\providecommand{\tightlist}{%
  \setlength{\itemsep}{0pt}\setlength{\parskip}{0pt}}
\usepackage{stata}
\usepackage{bm}
\usepackage[]{natbib}
\title{Simulating time-to-event data from parametric distributions, custom
distributions, competing risk models and general multi-state models}
\usepackage{etoolbox}
\makeatletter
\providecommand{\subtitle}[1]{
  \apptocmd{\@title}{\par {\large #1 \par}}{}{}
}
\makeatother
\subtitle{Michael J. Crowther}
\author{Karolinska Institutet}
\date{\href{mailto:michael.crowther@ki.se}{\nolinkurl{michael.crowther@ki.se}}}

\begin{document}
\maketitle
\begin{abstract}
In this paper I describe some substantial extensions to the
\texttt{survsim} command for simulating survival data. \texttt{survsim}
can now simulate survival data from a parametric distribution, a
custom/user-defined distribution, from a fitted \texttt{merlin} model,
from a specified cause-specific hazards competing risks model, or from a
specified general multi-state model (with multiple timescales). Left
truncation (delayed entry) is now also available for all settings. I
illustrate the command with some examples, demonstrating the huge
flexibilty that can be used to better evaluate statistical methods.
\end{abstract}

\begin{center}
\today
\end{center}

\hypertarget{introduction}{%
\section{Introduction}\label{introduction}}

\texttt{survsim} was first introduced with the ability to simulate from
a defined parametric distribution, including the exponential, Weibull,
Gompertz and 2-component versions of them. It allowed both
time-independent and time-dependent effects, and allowed simulation of
competing risks data, providing a useful tool to generate event times
\citep{Crowther2012b}. Following this, \citet{CrowtherSurvsim} developed
a general algorithm to simulate event times from a custom, user-defined
hazard or cumulative hazard function, using a combination of
root-finding and nested numerical integration, which was also
incorporated into \texttt{survsim}.

In this paper, I introduce some substantial new developments, including:

\begin{enumerate}
\def\labelenumi{\arabic{enumi}.}
\tightlist
\item
  The ability to simulate from a competing risks or general multi-state
  model. Event times can be simulated from transition-specific hazards,
  where each transition hazard function can be a standard parametric
  distribution, or a user-defined hazard function. Covariates and
  time-dependent effects can be specified for each transition-specific
  hazard independently, and multiple timescales can be incorporated. The
  simulation approach combines the competing risks simulation method
  described by \citet{Beyersmann2009}, and the general algorithm of
  \citet{CrowtherSurvsim}.
\item
  The ability to simulate from a conditional distribution, i.e.~allowing
  for delayed entry/left truncation, in all settings.
\item
  The ability to simulate directly from a fitted \texttt{merlin}
  survival model \citep{Crowther2020}. This is currently limited to
  univariate, observation-level only survival models, i.e.~no random
  effects.
\end{enumerate}

The paper is arranged as follows. In Section \ref{sec1} I briefly
describe the core algorithm used to simulate event times in
\texttt{survsim}. In Section \ref{sec:syntax}, I describe the syntax for
the four core settings that can be used to generate event times, which
are then illustrated in Section \ref{sec:eg}. I conclude in Section
\ref{sec:conc} with a brief discussion.

\hypertarget{simulating-survival-times}{%
\section{Simulating survival times}\label{simulating-survival-times}}

\label{sec1}

Simulating data from a defined distribution can be simple, or can be
extremely complex. Assume we have a random variable \(T\), with
associated cumulative distribution function, \(F(T)\). To simulate
survival times from such a distribution, we can simply let

\[
  \hspace{2ex} F \sim U(0,1)
\]

To simulate an observation, we draw from the uniform distribution, say
\(u \sim U(0,1)\), and simply substitute and solve for \(t\),

\[
F(t) = u
\]

and hence,

\[
t = F^{-1}(u)
\]

Now solving for \(t\) relies on being able to invert the cumulative
distribution function, and since it is a function of the cumulative
hazard function, we must be able to integrate our hazard function. To
accommodate these challenges, \citet{CrowtherSurvsim} developed a
combined root-finding and numerical integration algorithm to provide an
efficient method of generating event times from arbitrary distribution
functions. This forms the engine of the developments in this paper. For
the competing risks and more general multi-state setting, this is used
to simulate from the total hazard function (made up of the sum of
cause/transition-specific hazard functions leaving a particular state),
as developed by \citet{Beyersmann2009} for competing risks. Once a
survival time is simulated, a multinomial draw is conducted, with
probabilities equal to the cause/transition-specific hazards divided by
the total hazard, to decide which state is then entered. In the
multi-state setting, this is repeated as the observation transitions
from state to state, until either an absorbing state or a maximum
follow-up time is reached.

\hypertarget{the-survsim-command}{%
\section{\texorpdfstring{The \texttt{survsim}
command}{The survsim command}}\label{the-survsim-command}}

\label{sec:syntax}

Simulate survival data from a parametric distribution, a user-defined
distribution, from a fitted merlin model, from a cause-specific hazards
competing risks model, or from a general multi-state model.

\subsection{Syntax - simulate survival data from a parametric distribution}

  \vskip 0.25\baselineskip%
  \normalfont\ttfamily%
  \fontsize{10}{16}\selectfont%
  \hangindent=\stsyndent\hangafter=1\noindent%
  \raggedright%
  \hskip 0pt%

\hspace{2ex} survsim
    {\it newvarname\/}1
    {\tt $\bigl[$\,{{\it newvarname\/}2}\,$\bigr]$}
    ,
    distribution(string)
    {\tt $\bigl[$\,{,
    \texttt{\textit{options}}
    }\,$\bigr]$}
  \normalfont\selectfont%
  \vskip 0.25\baselineskip%
  \par%

where \textit{newvarname1} specifies the new variable name to contain
the generated survival times. \textit{newvarname2} is required when the
\texttt{maxtime()} option is specified which defines the time(s) of
right censoring.

\subsubsection{Options}

\hangindent=\parindent\hangafter=1\noindent

\texttt{noconstant} suppresses the constant (intercept) term and may be
specified for the fixed effects equation and for the random effects
equations.

\hangindent=\parindent\hangafter=1\noindent

\texttt{distribution(string)} specifies the parametric survival
distribution to use, including \texttt{exponential}, \texttt{gompertz}
or \texttt{weibull}. All listed distributions are parameterised in the
proportional hazards metric.

\hangindent=\parindent\hangafter=1\noindent

\texttt{lambdas(numlist)} defines the scale parameters in the
exponential/Weibull/Gompertz distribution(s). The number of values
required depends on the model choice. Default is a single number
corresponding to a standard parametric distribution. Under a mixture
model, 2 values are required.

\hangindent=\parindent\hangafter=1\noindent

\texttt{gammas(numlist)} defines the shape parameters of the
Weibull/Gompertz parametric distribution(s). Number of entries must be
equal to that of \texttt{lambdas()}.

\hangindent=\parindent\hangafter=1\noindent

\texttt{covariates(varname\ \#\ ...)} defines baseline covariates to be
included in the linear predictor of the survival model, along with the
value of the corresponding coefficient. For example, a treatment
variable coded 0/1 can be included, with a log hazard ratio of 0.5, by
\texttt{covariates(treat\ 0.5)}. Variable \texttt{treat} must be in the
dataset before \texttt{survsim} is called.

\hangindent=\parindent\hangafter=1\noindent

\texttt{tde(varname\ \#\ ...)} creates non-proportional hazards by
interacting covariates with log time for an exponential or Weibull
model, or time under a Gompertz model or mixture model. Values should be
entered as \texttt{tde(trt\ 0.5)}, for example. Multiple time-dependent
effects can be specified, but they will all be interacted with the same
function of time.

\hangindent=\parindent\hangafter=1\noindent

\texttt{maxtime(\#\textbar{}varname)} specifies the right censoring
time(s). Either a common maximum follow-up time \texttt{\#} can be
specified for all observations, or observation specific censoring times
can be specified by using a \texttt{varname}.

\hangindent=\parindent\hangafter=1\noindent

\texttt{ltruncated(\#\textbar{}varname)} specifies the left
truncated/delayed entry time(s), to simulate from a conditional survival
distribution. Either a common time \texttt{\#} can be specified for all
observations, or observation specific left truncation times can be
specified by using a \texttt{varname}.

\hangindent=\parindent\hangafter=1\noindent

\texttt{mixture} specifies that survival times are simulated from a
2-component mixture model, with mixture component distributions defined
by \texttt{distribution()}. \texttt{lambdas()} and \texttt{gammas()}
must be of length 2.

\hangindent=\parindent\hangafter=1\noindent

\texttt{pmix(\#)} defines the mixture parameter. Default is 0.5.

\hangindent=\parindent\hangafter=1\noindent

\texttt{nodes(\#)} defines the number of Gauss-Legendre quadrature
points used to evaluate the cumulative hazard function when
\texttt{mixture} and \texttt{tde()} are specified together. To simulate
survival times from such a mixture model, a combination of numerical
integration and root-finding is used. The default is \texttt{nodes(30)}.

\subsection{Syntax for simulating survival times from a user-defined distribution}

  \vskip 0.25\baselineskip%
  \normalfont\ttfamily%
  \fontsize{10}{16}\selectfont%
  \hangindent=\stsyndent\hangafter=1\noindent%
  \raggedright%
  \hskip 0pt%

\hspace{2ex} survsim
    {\it newvarname\/}1
    {\it newvarname\/}2
    ,
    maxtime({\it \#\/}|varname)
    {\tt $\bigl[$\,{,
    \texttt{\textit{options}}
    }\,$\bigr]$}
  \normalfont\selectfont%
  \vskip 0.25\baselineskip%
  \par%

where \textit{newvarname1} specifies the new variable name to contain
the generated survival times, and \textit{newvarname2} specifies the new
variable name to contain the generated event indicator.

\subsubsection{Options}

\hangindent=\parindent\hangafter=1\noindent

\texttt{loghazard(string)} is the user-defined log hazard function. The
function can include:

\begin{itemize}
\tightlist
\item
  \texttt{\{t\}} which denotes the main timescale. If
  \texttt{ltrunacted()} is specified, it is also measured on this
  timescale.
\item
  \texttt{varname} which denotes a variable in your dataset
\item
  \texttt{+-*/\^{}} standard Mata mathematical operators, using colon
  notation i.e.~\texttt{2\ :*\ \{t\}}, see
  \texttt{help\ {[}M-2{]}\ op\_colon}. Colon operators must be used as
  behind the scenes, \texttt{\{t\}} gets replaced by an \texttt{\_N} by
  \texttt{nodes()} matrix when numerically integrating the hazard
  function.
\item
  \texttt{mata\_function()} any Mata function, e.g.~\texttt{log()} and
  \texttt{exp()}
\end{itemize}

\hangindent=\parindent\hangafter=1\noindent

\texttt{hazard(string)} is the user-defined baseline hazard function.
See \texttt{loghazard()} for more details, and examples below.

\hangindent=\parindent\hangafter=1\noindent

\texttt{logchazard(string)} is the user-defined log cumulative baseline
hazard function. See \texttt{loghazard()} for more details, and examples
below.

\hangindent=\parindent\hangafter=1\noindent

\texttt{chazard(string)} is the user-defined baseline cumulative hazard
function. See \texttt{loghazard()} for more details, and examples below.

\hangindent=\parindent\hangafter=1\noindent

\texttt{covariates(varname\ \#\ ...)} defines baseline covariates to be
included in the linear predictor of the survival model, along with the
value of the corresponding coefficient. For example, a treatment
variable coded 0/1 can be included, with a log hazard ratio of 0.5, by
\texttt{covariates(treat\ 0.5)}. Variable treat must be in the dataset
before survsim is called. If \texttt{chazard()} or \texttt{logchazard()}
are used, then \texttt{covariates()} effects are additive on the log
cumulative hazard scale.

\hangindent=\parindent\hangafter=1\noindent

\texttt{tde(varname\ \#\ ...)} creates non-proportional hazards by
interacting covariates with a function of time, defined by
tdefunction(), on the appropriate log hazard or log cumulative hazard
scale. Values should be entered as \texttt{tde(trt\ 0.5)}, for example.
Multiple time-dependent effects can be specified, but they will all be
interacted with the same \texttt{tdefunction()}. To circumvent this, you
can directly specify them in your user function.

\hangindent=\parindent\hangafter=1\noindent

\texttt{tdefunction(string)} defines the function of time to which
covariates specified in \texttt{tde()} are interacted with, to create
time-dependent effects. The default is \{t\}, i.e.~linear time. The
function can include:

\begin{itemize}
\tightlist
\item
  \texttt{\{t\}} which denotes the main timescale, measured on the time
  since starting state, \texttt{startstate()}, timescale (which may be
  \texttt{ltruncated()})
\item
  \texttt{+-*/\^{}} standard Mata mathematical operators, using colon
  notation i.e.~\texttt{2\ :*\ \{t\}}, see
  \texttt{help\ {[}M-2{]}\ op\_colon}. Colon operators must be used as
  behind the scenes, \texttt{\{t\}} gets replaced by an \_N x nodes()
  matrix when numerically integrating the hazard function.
  mata\_function any Mata function, e.g.~log() and exp()
\end{itemize}

\hangindent=\parindent\hangafter=1\noindent

\texttt{maxtime(\#\textbar{}varname)} specifies the right censoring
time(s). Either a common maximum follow-up time \# can be specified for
all observations, or observation specific censoring times can be
specified by using a \texttt{varname}.

\hangindent=\parindent\hangafter=1\noindent

\texttt{ltruncated(\#\textbar{}varname)} specifies the left
truncated/delayed entry time(s). Either a common time \texttt{\#} can be
specified for all observations, or observation specific left truncation
times can be specified by using a \texttt{varname}.

\hangindent=\parindent\hangafter=1\noindent

\texttt{nodes(\#)} defines the number of Gauss-Legendre quadrature
points used to evaluate the cumulative hazard function when
\texttt{loghazard()} or \texttt{hazard()} is specified. To simulate
survival times from such a function, a combination of numerical
integration and root-finding is used. The default is \texttt{nodes(30)}.

\subsection{Syntax for simulating survival times from a fitted merlin survival model}

  \vskip 0.25\baselineskip%
  \normalfont\ttfamily%
  \fontsize{10}{16}\selectfont%
  \hangindent=\stsyndent\hangafter=1\noindent%
  \raggedright%
  \hskip 0pt%

\hspace{2ex} survsim
    {\it newvarname\/}1
    {\it newvarname\/}2
    ,
    model(name)
    maxtime({\it \#\/}|varname)
  \normalfont\selectfont%
  \vskip 0.25\baselineskip%
  \par%

where \textit{newvarname1} specifies the new variable name to contain
the generated survival times, and \textit{newvarname2} specifies the new
variable name to contain the generated event indicator.

\subsubsection{Options}

\hangindent=\parindent\hangafter=1\noindent

\texttt{model(name)} specifies the name of the \texttt{estimates\ store}
object containing the estimates of the model fitted. The survival model
must be estimated using the \texttt{merlin} command. \texttt{survsim}
will simulate from the fitted model, using covariate values that are in
your current dataset. For example,

\begin{verbatim}
    . merlin (_t trt , family(weibull, failure(_d)))
    . estimates store m1
    . survsim stime died, model(m1) maxtime(10)
\end{verbatim}

\hangindent=\parindent\hangafter=1\noindent

\texttt{maxtime(\#\textbar{}varname)} specifies the right censoring
time(s). Either a common maximum follow-up time \texttt{\#} can be
specified for all observations, or observation specific censoring times
can be specified by using a \texttt{varname}.

\subsection{Syntax for simulating survival times from a multi-state model}

  \vskip 0.25\baselineskip%
  \normalfont\ttfamily%
  \fontsize{10}{16}\selectfont%
  \hangindent=\stsyndent\hangafter=1\noindent%
  \raggedright%
  \hskip 0pt%

\hspace{2ex} survsim
    \textit{timestub}
    \textit{statestub}
    \textit{eventstub}
    ,
    hazard1(haz\_options)
    hazard2(haz\_options)
    maxtime({\it \#\/}|varname)
    {\tt $\bigl[$\,{,
        transmatrix(name)
        hazard3(haz\_options)
        \texttt{\textit{options}}
    }\,$\bigr]$}
  \normalfont\selectfont%
  \vskip 0.25\baselineskip%
  \par%

where \textit{timestub} specifies the new variable name stub to contain
the generated survival times for each transition. \textit{statestub}
specifies the new variable name stub to contain the generated state
identifiers, i.e which state each observation has transitioned to/is in,
at the associated times. \textit{eventstub} specifies the new variable
name stub to contain the event indicators, i.e.~whether an event or
right-censoring occurred at the associated times. When observations have
reached an absorbing state, or reached \texttt{maxtime()}, they will
have missing observations in any further generated variables. When all
observations have reached an absorbing state, or are right-censored,
then the simulation will cease and the generated variables are returned.

\subsubsection{Options}

\hangindent=\parindent\hangafter=1\noindent

\texttt{distribution(string)} specifies the parametric survival
distribution to use, including \texttt{exponential}, \texttt{gompertz}
or \texttt{weibull}. All listed distributions are parameterised in the
proportional hazards metric.

\hangindent=\parindent\hangafter=1\noindent

\texttt{lambda(\#)} defines the scale parameter for a
exponential/Weibull/Gompertz distribution.

\hangindent=\parindent\hangafter=1\noindent

\texttt{gamma(\#)} defines the shape parameter for a Weibull/Gompertz
parametric distribution.

\hangindent=\parindent\hangafter=1\noindent

\texttt{user(function)} defines a custom hazard function. The function
can include:

\begin{itemize}
\tightlist
\item
  \texttt{\{t\}} which denotes the main timescale. If
  \texttt{ltruncated()} is specified, it is also measured on this
  timescale.
\item
  \texttt{\{t0\}} which denotes the time of entry to the current state
  of the associated transition hazard, measured on the time since
  initial starting state timescale
\item
  \texttt{varname} which denotes a variable in your dataset
\item
  \texttt{+-*/\^{}} standard Mata mathematical operators, using colon
  notation i.e.~\texttt{2\ :*\ \{t\}}, see
  \texttt{help\ {[}M-2{]}\ op\_colon}. Colon operators must be used as
  behind the scenes, \texttt{\{t\}} gets replaced by an \texttt{\_N} by
  \texttt{nodes()} matrix when numerically integrating the transition
  hazard function.
\item
  \texttt{mata\_function} any Mata function, e.g.~\texttt{log()} and
  \texttt{exp()}
\end{itemize}

For example,

\begin{verbatim}
    dist(weibull) lambda(0.1) gamma(1.2)
\end{verbatim}

is equivalent to

\begin{verbatim}
    user(0.1:*1.2:*{t}:^(1.2:-1))
\end{verbatim}

\hangindent=\parindent\hangafter=1\noindent

\texttt{covariates(varname\ \#\ ...)} defines baseline covariates to be
included in the linear predictor of the transition-specific hazard
function, along with the value of the corresponding coefficient. For
example, a treatment variable coded 0/1 can be included, with a log
hazard ratio of 0.5, by \texttt{covariates(treat\ 0.5)}. Variable treat
must be in the dataset before \texttt{survsim} is called.

\hangindent=\parindent\hangafter=1\noindent

\texttt{tde(varname\ \#\ ...)} creates non-proportional hazards by
interacting covariates with a function of time. Covariates are
interacted with \texttt{tdefunction()}, on the log hazard scale. Values
should be entered as \texttt{tde(trt\ 0.5)}, for example. Multiple
time-dependent effects can be specified, but they will all be interacted
with the same function of time.

\hangindent=\parindent\hangafter=1\noindent

\texttt{tdefunction(string)} defines the function of time to which
covariates specified in \texttt{tde()} are interacted with, to create
time-dependent effects in the transition-specific hazard function. The
default is \texttt{\{t\}}, i.e.~linear time. The function can include:

\begin{itemize}
\tightlist
\item
  \texttt{\{t\}} which denotes the main timescale
\item
  \texttt{+-*/\^{}} standard Mata mathematical operators, using colon
  notation i.e.~\texttt{2\ :*\ \{t\}}, see
  \texttt{help\ {[}M-2{]}\ op\_colon}. Colon operators must be used as
  behind the scenes, \texttt{\{t\}} gets replaced by an \texttt{\_N} by
  \texttt{nodes()} matrix when numerically integrating the transition
  hazard function.
\item
  \texttt{mata\_function} any Mata function, e.g.~\texttt{log()} and
  \texttt{exp()}
\end{itemize}

\hangindent=\parindent\hangafter=1\noindent

\texttt{reset} specifies that this transition model is on a clock-reset
timescale. The timescale is reset to 0 on entry, i.e.~the timescale for
this transition is measured on a time since state entry timescale,
rather than the default clock-forward. If you specify a \texttt{user()}
function with reset, then survsim will replace any occurrences of
\texttt{\{t\}} with \texttt{\{t\}-\{t0\}}, including those specified in
\texttt{tdefunction()}.

\hangindent=\parindent\hangafter=1\noindent

\texttt{transmatrix(matname)} specifies the transition matrix which
governs the multi-state model. Transitions must be numbered as an
increasing sequence of integers from 1,\ldots,K, from left to right, top
to bottom of the matrix. Reversible transitions are allowed. If
\texttt{transmatrix()} is not specified, a competing risks model is
assumed.

\hangindent=\parindent\hangafter=1\noindent

\texttt{maxtime(\#\textbar{}varname)} specifies right censoring time(s).
Either a common maximum follow-up time \texttt{\#} can be specified for
all observations, or observation specific censoring times can be
specified by using a \texttt{varname}.

\hangindent=\parindent\hangafter=1\noindent

\texttt{startstate(\#\textbar{}varname)} specifies the state(s) in which
observations begin. Either a common state \texttt{\#} can be specified
for all observations, or observation specific starting states can be
specified by using a \texttt{varname}. Default is
\texttt{startstate(1)}.

\hangindent=\parindent\hangafter=1\noindent

\texttt{ltruncated(\#\textbar{}varname)} specifies left
truncated/delayed entry time(s), which is the time(s) at which
observations start in the initial starting state(s). Either a common
time \texttt{\#} can be specified for all observations, or observation
specific left truncation times can be specified by using a varname.
Default is \texttt{ltruncated(0)}.

\hangindent=\parindent\hangafter=1\noindent

\texttt{nodes(\#)} defines the number of Gauss-Legendre quadrature
points used to evaluate the total cumulative hazard function for each
potential next transition. To simulate survival times from such a
function, a combination of numerical integration and root-finding is
used. The default is \texttt{nodes(30)}.

\section{Examples}
\label{sec:eg}

In this section I will go through at least one example of how to use
\texttt{survsim} to simulate survival data from each of the four main
settings.

\subsection{Simulating survival times from standard parametric distributions}

Let's simulate survival times from a Weibull distribution, with a binary
treatment group, \texttt{trt}, and a continuous covariate, \texttt{age},
under proportional hazards:

\[
h(t) = \lambda \gamma t^{\gamma - 1} \exp (\text{trt} \beta_{1} + \text{age} \beta_{2})
\]

I'll simulate 300 observations, and pick some distributions for the
covariates, which should be self-explanatory,

\begin{Shaded}
\begin{Highlighting}[]
\NormalTok{. }\KeywordTok{clear}

\NormalTok{. }\KeywordTok{set} \KeywordTok{obs}\NormalTok{ 300}
\NormalTok{Number }\KeywordTok{of}\NormalTok{ observations (_N) was 0, now 300.}

\NormalTok{. }\KeywordTok{set} \DecValTok{seed}\NormalTok{ 134987}

\NormalTok{. }\KeywordTok{gen}\NormalTok{ trt = runiform()>0.5}

\NormalTok{. }\KeywordTok{gen}\NormalTok{ age = rnormal(50,3)}
\end{Highlighting}
\end{Shaded}

We then call \texttt{survsim}, setting \(\lambda = 0.1\),
\(\gamma=1.2\), \(\beta_{1}=-0.5\) and \(\beta_{2}=0.01\),

\begin{Shaded}
\begin{Highlighting}[]
\NormalTok{. survsim stime, distribution(}\KeywordTok{weibull}\NormalTok{) lambda(0.1) }\KeywordTok{gamma}\NormalTok{(1.2)  }\CommentTok{///}
\NormalTok{>                covariates(trt -0.5 age 0.01)}
\end{Highlighting}
\end{Shaded}

which stores our simulated survival times in the new variable
\texttt{stime}. If we wanted to apply right-censoring, we could first
generate some censoring times, we could apply a common censoring time
using for example \texttt{maxtime(5)}, which would censor all
observations if their simulated event times was greater than 5, or we
could generate observation specific potential censoring times, such as

\begin{Shaded}
\begin{Highlighting}[]
\NormalTok{. }\KeywordTok{gen}\NormalTok{ censtime = runiform() * 5}
\end{Highlighting}
\end{Shaded}

and now add the \texttt{maxtime()} option to \texttt{survsim},
remembering to also specify a second new variable name for the event
indicator,

\begin{Shaded}
\begin{Highlighting}[]
\NormalTok{. survsim stime2 died2, distribution(}\KeywordTok{weibull}\NormalTok{) lambda(0.1) }\KeywordTok{gamma}\NormalTok{(1.2)     }\CommentTok{///}
\NormalTok{>                       covariates(trt -0.5 age 0.01) maxtime(censtime)}
\end{Highlighting}
\end{Shaded}

We could also:

\begin{itemize}
\tightlist
\item
  add time-dependent effects using the \texttt{tde()} option
\item
  add left-truncation/delayed entry using the \texttt{ltruncated()}
  option
\item
  simulate from a 2-component mixture distribution using the
  \texttt{mixture} option
\end{itemize}

\subsection{Simulating survival times from a user-defined (log) (cumulative) hazard function}

This section illustrates how to simulate from a user-defined function,
first described in \citet{CrowtherSurvsim}. Let's start by simulating
500 observations, and generate a binary treatment group,

\begin{Shaded}
\begin{Highlighting}[]
\NormalTok{. }\KeywordTok{clear}

\NormalTok{. }\KeywordTok{set} \KeywordTok{obs}\NormalTok{ 500}
\NormalTok{Number }\KeywordTok{of}\NormalTok{ observations (_N) was 0, now 500.}

\NormalTok{. }\KeywordTok{set} \DecValTok{seed}\NormalTok{ 134987}

\NormalTok{. }\KeywordTok{gen}\NormalTok{ trt = runiform()>0.5}
\end{Highlighting}
\end{Shaded}

The most flexible form of simulating survival data with \texttt{survsim}
is by specifying a custom hazard or cumulative hazard function, such as:

\[
h(t) = h_{0}(t) \exp (\text{trt} \beta_{1})
\] where \[
h_{0}(t) = \exp (-1 + 0.02 t - 0.03 t^2 + 0.005 t^3)
\]

which can be done, on the \texttt{loghazard()} scale for simplicity,
using

\begin{Shaded}
\begin{Highlighting}[]
\NormalTok{. survsim stime1 died1, loghazard(-1:+0.02:*\{t\}:-0.03:*\{t\}:^2:+0.005:*\{t\}:^3)  }\CommentTok{///}
\NormalTok{>                       covariates(trt -0.5) maxtime(1.5)}
\NormalTok{Warning: 321 survival times were above the }\FunctionTok{upper}\NormalTok{ limit }\KeywordTok{of}\NormalTok{ maxtime()}
\NormalTok{         They have been }\KeywordTok{set}\NormalTok{ to maxtime()}
\NormalTok{         You can identify them }\KeywordTok{by}\NormalTok{ _survsim_rc = 3}
\end{Highlighting}
\end{Shaded}

The \texttt{loghazard()} function is defined using Mata code, with colon
operators representing element by element operations. Time must be
referred to using the \texttt{\{t\}} notation. A common right censoring
time of 1.5 years is specified using \texttt{maxtime(1.5)}. We could
make the treatment effect diminish over log time by incorporating a
time-dependent effect, where

\[
\beta_{1}(t) = \log(t) \beta_{1}
\]

which is defined using the \texttt{tdefunction()} and \texttt{tde()}
options, setting \(\beta_{1}=0.03\)

\begin{Shaded}
\begin{Highlighting}[]
\NormalTok{. survsim stime2 died2, loghazard(-1:+0.02:*\{t\}:-0.03:*\{t\}:^2:+0.005:*\{t\}:^3) }\CommentTok{///}
\NormalTok{>                       covariates(trt -0.5) tde(trt 0.03) tdefunc(}\FunctionTok{log}\NormalTok{(\{t\}))  }\CommentTok{///}
\NormalTok{>                       maxtime(1.5)}
\NormalTok{Warning: 328 survival times were above the }\FunctionTok{upper}\NormalTok{ limit }\KeywordTok{of}\NormalTok{ maxtime()}
\NormalTok{         They have been }\KeywordTok{set}\NormalTok{ to maxtime()}
\NormalTok{         You can identify them }\KeywordTok{by}\NormalTok{ _survsim_rc = 3}
\end{Highlighting}
\end{Shaded}

which will form an interaction between \texttt{trt}, its coefficient
\texttt{0.03} and log time. Alternatively, we could instead simulate
from a model on the cumulative hazard scale, using the
\texttt{logchazard()} option instead.

\subsection{Simulating survival times from a fitted \texttt{merlin} survival model}

Rather than simulating from a particular data-generating model specified
essentially by hand, we can directly simulate from a fitted model, by
passing an \texttt{estimates} object to \texttt{survsim} through the
\texttt{model()} option. This is similar to \texttt{stsurvsim}
\citep{Roystonstsurvsim} which allows the simulation of survival times
from a Royston-Parmar flexible parametric model. \texttt{survsim} now
allows the simulation from survival model that has been fitted with the
\texttt{merlin} command. Let's fit a Weibull model to a standard
survival dataset:

\begin{Shaded}
\begin{Highlighting}[]
\NormalTok{. }\KeywordTok{webuse}\NormalTok{ brcancer, }\KeywordTok{clear}
\NormalTok{(German breast cancer }\KeywordTok{data}\NormalTok{)}

\NormalTok{. }\KeywordTok{stset}\NormalTok{ rectime, f(censrec=1) }\KeywordTok{scale}\NormalTok{(365)}

\NormalTok{Survival-time }\KeywordTok{data}\NormalTok{ settings}

\NormalTok{         Failure event: censrec==1}
\NormalTok{Observed time interval: (0, rectime]}
\NormalTok{     Exit }\KeywordTok{on} \KeywordTok{or}\NormalTok{ before: failure}
\NormalTok{     Time }\KeywordTok{for}\NormalTok{ analysis: time/365}

\NormalTok{--------------------------------------------------------------------------}
\NormalTok{        686  }\KeywordTok{total}\NormalTok{ observations}
\NormalTok{          0  exclusions}
\NormalTok{--------------------------------------------------------------------------}
\NormalTok{        686  observations remaining, representing}
\NormalTok{        299  failures }\KeywordTok{in}\NormalTok{ single-record/single-failure }\KeywordTok{data}
\NormalTok{  2,113.425  }\KeywordTok{total}\NormalTok{ analysis time }\FunctionTok{at}\NormalTok{ risk and under observation}
\NormalTok{                                                At risk from t =         0}
\NormalTok{                                     Earliest observed entry t =         0}
\NormalTok{                                          Last observed }\KeywordTok{exit}\NormalTok{ t =  7.284932}

\NormalTok{. merlin (_t hormon , }\KeywordTok{family}\NormalTok{(}\KeywordTok{weibull}\NormalTok{, failure(_d)))}

\NormalTok{Fitting full }\KeywordTok{model}\NormalTok{:}

\NormalTok{Iteration 0:   }\FunctionTok{log}\NormalTok{ likelihood = -2305.7955  }
\NormalTok{Iteration 1:   }\FunctionTok{log}\NormalTok{ likelihood = -901.52031  }
\NormalTok{Iteration 2:   }\FunctionTok{log}\NormalTok{ likelihood = -870.50824  }
\NormalTok{Iteration 3:   }\FunctionTok{log}\NormalTok{ likelihood = -868.05956  }
\NormalTok{Iteration 4:   }\FunctionTok{log}\NormalTok{ likelihood = -868.02684  }
\NormalTok{Iteration 5:   }\FunctionTok{log}\NormalTok{ likelihood = -868.02684  }

\NormalTok{Fixed }\KeywordTok{effects}\NormalTok{ regression }\KeywordTok{model}\NormalTok{                             Number }\KeywordTok{of} \KeywordTok{obs}\NormalTok{ = 686}
\NormalTok{Log likelihood = -868.02684}
\NormalTok{------------------------------------------------------------------------------}
\NormalTok{             | Coefficient  Std. err.      z    P>|z|     [95% conf. interval]}
\NormalTok{-------------+----------------------------------------------------------------}
\NormalTok{_t:          |            }
\NormalTok{      hormon |  -.3932403   .1248267    -3.15   0.002    -.6378961   -.1485845}
       \DataTypeTok{_cons}\NormalTok{ |  -2.196011   .1094092   -20.07   0.000     -2.41045   -1.981573}
  \FunctionTok{log}\NormalTok{(}\KeywordTok{gamma}\NormalTok{) |    .250997   .0496958     5.05   0.000     .1535949     .348399}
\NormalTok{------------------------------------------------------------------------------}
\end{Highlighting}
\end{Shaded}

I'm using the benefits of \texttt{stset} to declare the survival
variables, which I can use directly within \texttt{merlin} for
simplicity. We then simply store the model object, calling it whatever
we like, such as the imaginative name of \texttt{m1},

\begin{Shaded}
\begin{Highlighting}[]
\NormalTok{. }\KeywordTok{estimates} \KeywordTok{store}\NormalTok{ m1}
\end{Highlighting}
\end{Shaded}

This we then pass to \texttt{survsim} to simulate a dataset, of the same
size, using our fitted results,

\begin{Shaded}
\begin{Highlighting}[]
\NormalTok{. survsim stime5 died5, }\KeywordTok{model}\NormalTok{(m1) maxtime(7)}
\end{Highlighting}
\end{Shaded}

The option \texttt{maxtime()} is required in this case. We can then fit
the same model as before, and of course get slightly different results,
because we have sampling variability.

\begin{Shaded}
\begin{Highlighting}[]
\NormalTok{. }\KeywordTok{stset}\NormalTok{ stime5, failure(died5)}

\NormalTok{Survival-time }\KeywordTok{data}\NormalTok{ settings}

\NormalTok{         Failure event: died5!=0 & died5<.}
\NormalTok{Observed time interval: (0, stime5]}
\NormalTok{     Exit }\KeywordTok{on} \KeywordTok{or}\NormalTok{ before: failure}

\NormalTok{--------------------------------------------------------------------------}
\NormalTok{        686  }\KeywordTok{total}\NormalTok{ observations}
\NormalTok{          0  exclusions}
\NormalTok{--------------------------------------------------------------------------}
\NormalTok{        686  observations remaining, representing}
\NormalTok{        447  failures }\KeywordTok{in}\NormalTok{ single-record/single-failure }\KeywordTok{data}
\NormalTok{  3,166.854  }\KeywordTok{total}\NormalTok{ analysis time }\FunctionTok{at}\NormalTok{ risk and under observation}
\NormalTok{                                                At risk from t =         0}
\NormalTok{                                     Earliest observed entry t =         0}
\NormalTok{                                          Last observed }\KeywordTok{exit}\NormalTok{ t =         7}

\NormalTok{. merlin (_t hormon , }\KeywordTok{family}\NormalTok{(}\KeywordTok{weibull}\NormalTok{, failure(_d)))}

\NormalTok{Fitting full }\KeywordTok{model}\NormalTok{:}

\NormalTok{Iteration 0:   }\FunctionTok{log}\NormalTok{ likelihood = -3455.2148  }
\NormalTok{Iteration 1:   }\FunctionTok{log}\NormalTok{ likelihood = -1350.3894  }
\NormalTok{Iteration 2:   }\FunctionTok{log}\NormalTok{ likelihood = -1315.2054  }
\NormalTok{Iteration 3:   }\FunctionTok{log}\NormalTok{ likelihood = -1299.6613  }
\NormalTok{Iteration 4:   }\FunctionTok{log}\NormalTok{ likelihood = -1298.8091  }
\NormalTok{Iteration 5:   }\FunctionTok{log}\NormalTok{ likelihood =  -1298.806  }
\NormalTok{Iteration 6:   }\FunctionTok{log}\NormalTok{ likelihood =  -1298.806  }

\NormalTok{Fixed }\KeywordTok{effects}\NormalTok{ regression }\KeywordTok{model}\NormalTok{                             Number }\KeywordTok{of} \KeywordTok{obs}\NormalTok{ = 686}
\NormalTok{Log likelihood = -1298.806}
\NormalTok{------------------------------------------------------------------------------}
\NormalTok{             | Coefficient  Std. err.      z    P>|z|     [95% conf. interval]}
\NormalTok{-------------+----------------------------------------------------------------}
\NormalTok{_t:          |            }
\NormalTok{      hormon |  -.3533906   .1014646    -3.48   0.000    -.5522575   -.1545237}
       \DataTypeTok{_cons}\NormalTok{ |  -2.349859   .1100583   -21.35   0.000    -2.565569   -2.134149}
  \FunctionTok{log}\NormalTok{(}\KeywordTok{gamma}\NormalTok{) |   .2650626   .0421563     6.29   0.000     .1824378    .3476875}
\NormalTok{------------------------------------------------------------------------------}
\end{Highlighting}
\end{Shaded}

Note, \texttt{survsim} will simulate the same number of observations as
\texttt{\_N}, so any covariates in your model must have non-missing
observations, otherwise missing values will be produced. The useful
aspect of this, is that you can manipulate your covariate distributions
in your dataset, and then simply recall \texttt{survsim} to simulate new
survival times using the same estimated parameters.

Currently, \texttt{survsim} can simulate from any of the available
survival models in \texttt{merlin}, but does not support simulation from
multivariate models or a model containing random effects.

\subsection{Simulating competing risks data from specified cause-specific hazard functions}

\texttt{survsim} simulates competing risk data using the cause-specific
hazard setting described by \citet{Beyersmann2009}, utilising the
general simulation algorithm described by \citet{CrowtherSurvsim}. Let's
simulate from a competing risk model with 2 competing events. The first
cause-specific hazard has a Weibull distribution, with no covariates.
The second cause-specific hazard model has an exponential distribution,
with a beneficial treatment effect. Right censoring is applied at 10
years.

\begin{Shaded}
\begin{Highlighting}[]
\NormalTok{. }\KeywordTok{clear}

\NormalTok{. }\KeywordTok{set} \DecValTok{seed}\NormalTok{ 398}

\NormalTok{. }\KeywordTok{set} \KeywordTok{obs}\NormalTok{ 1000}
\NormalTok{Number }\KeywordTok{of}\NormalTok{ observations (_N) was 0, now 1,000.}

\NormalTok{. }\KeywordTok{gen}\NormalTok{ trt = runiform()>0.5}

\NormalTok{. survsim time state event , hazard1(dist(}\KeywordTok{weibull}\NormalTok{) lambda(0.1) }\KeywordTok{gamma}\NormalTok{(0.8))  }\CommentTok{///}
\NormalTok{>                            hazard2(dist(exponential) lambda(0.02)         }\CommentTok{/// }
\NormalTok{>                            covariates(trt -0.5)) maxtime(10)}
\NormalTok{variables time0 to time1 created}
\NormalTok{variables state0 to state1 created}
\NormalTok{variables event1 to event1 created}
\end{Highlighting}
\end{Shaded}

Each \texttt{hazard\#()} defines a cause-specific hazard function, with
specified \texttt{distribution()} and associated baseline parameters,
covariate effects and time-dependent effects. Each of the
\texttt{hazard\#()} functions can be as similar, or different, as
required. \texttt{survsim} creates some new variables, based on the
\textit{newvarstubs} that we specified in the call.

\begin{Shaded}
\begin{Highlighting}[]
\NormalTok{. }\OtherTok{list} \KeywordTok{if} \DataTypeTok{_n}\NormalTok{<=5}

\NormalTok{      +----------------------------------------------------+}
\NormalTok{      | trt   time0   state0       time1   state1   event1 |}
\NormalTok{      |----------------------------------------------------|}
\NormalTok{   1. |   0       0        1   4.5792847        2        1 |}
\NormalTok{   2. |   1       0        1          10        1        0 |}
\NormalTok{   3. |   1       0        1          10        1        0 |}
\NormalTok{   4. |   1       0        1   2.8415219        3        1 |}
\NormalTok{   5. |   0       0        1    1.576534        2        1 |}
\NormalTok{      +----------------------------------------------------+}
\end{Highlighting}
\end{Shaded}

Since the competing risks simulation is framed within a more general
multi-state setting, discussed in the next example, all observations are
assumed to begin in an initial starting state 1, at an initial starting
time 0. These are stored in \texttt{state0} and \texttt{time0},
respectively. The starting time can be changed using the
\texttt{ltruncated()} option.

From the starting state, observations have two places to go:

\begin{itemize}
\tightlist
\item
  State 1 to State 2, with the transition rate governed by
  \texttt{hazard1()}
\item
  State 1 to State 3, with the transition rate governed by
  \texttt{hazard2()}
\end{itemize}

We can see which events occurred with

\begin{Shaded}
\begin{Highlighting}[]
\NormalTok{. }\KeywordTok{tabulate}\NormalTok{ state1 event1}

\NormalTok{           |        event1}
\NormalTok{    state1 |         0          1 |     Total}
\NormalTok{-----------+----------------------+----------}
\NormalTok{         1 |       484          0 |       484 }
\NormalTok{         2 |         0        414 |       414 }
\NormalTok{         3 |         0        102 |       102 }
\NormalTok{-----------+----------------------+----------}
\NormalTok{     Total |       484        516 |     1,000 }
\end{Highlighting}
\end{Shaded}

which shows that at by ten years, 484 observations were right-censored,
414 are in State 2, and 102 are in State 3.

Now let's simulate from a competing risk model with 3 competing events.
The first cause-specific hazard has a user defined baseline hazard
function, with a harmful treatment effect. The second cause-specific
hazard model has a Weibull distribution, with a beneficial treatment
effect. The third cause-specific hazard has a user-defined baseline
hazard function, with an initially beneficial treatment effect that
reduces linearly with respect to log time. Right censoring is applied at
3 years. Phew.

\begin{Shaded}
\begin{Highlighting}[]
\NormalTok{. cap }\KeywordTok{drop}\NormalTok{ time* state* event*}

\NormalTok{. }\KeywordTok{set} \DecValTok{seed}\NormalTok{ 32984575}

\NormalTok{. survsim time state event,                                    }\CommentTok{///}
\NormalTok{>         hazard1(user(}\FunctionTok{exp}\NormalTok{(-2 :+ 0.2:* }\FunctionTok{log}\NormalTok{(\{t\}) :+ 0.1:*\{t\}))  }\CommentTok{///}
\NormalTok{>                 covariates(trt 0.1))                         }\CommentTok{///}
\NormalTok{>         hazard2(dist(}\KeywordTok{weibull}\NormalTok{) lambda(0.01) }\KeywordTok{gamma}\NormalTok{(1.3)        }\CommentTok{///}
\NormalTok{>                 covariates(trt -0.5))                        }\CommentTok{///}
\NormalTok{>         hazard3(user(0.1 :* \{t\} :^ 1.5) covariates(trt -0.5) }\CommentTok{///}
\NormalTok{>                 tde(trt 0.1) tdefunction(}\FunctionTok{log}\NormalTok{(\{t\})))          }\CommentTok{///}
\NormalTok{>         maxtime(3)}
\NormalTok{variables time0 to time1 created}
\NormalTok{variables state0 to state1 created}
\NormalTok{variables event1 to event1 created}

\NormalTok{. }\KeywordTok{tabulate}\NormalTok{ state1 event1}

\NormalTok{           |        event1}
\NormalTok{    state1 |         0          1 |     Total}
\NormalTok{-----------+----------------------+----------}
\NormalTok{         1 |       341          0 |       341 }
\NormalTok{         2 |         0        345 |       345 }
\NormalTok{         3 |         0         30 |        30 }
\NormalTok{         4 |         0        284 |       284 }
\NormalTok{-----------+----------------------+----------}
\NormalTok{     Total |       341        659 |     1,000 }
\end{Highlighting}
\end{Shaded}

You can see that this can get as complex as necessary. I currently let
you use up to 50 cause-specific hazards, just in case you're feeling
particularly adventurous.

\subsection{Simulating from an illness-death model}
\label{sec:illdeg}

We first define the transition matrix for an illness-death model. It has
three states:

\begin{itemize}
\tightlist
\item
  State 1 - A ``healthy'' state. Observations can move from state 1 to
  state 2 or 3.
\item
  State 2 - An intermediate ``illness'' state. Observations can come
  from state 1, and move on to state 3.
\item
  State 3 - An absorbing ``death'' state. Observations can come from
  state 1 or 2, but not leave.
\end{itemize}

This gives us three potential transitions between states:

\begin{itemize}
\tightlist
\item
  Transition 1 - State 1 -\textgreater{} State 2
\item
  Transition 2 - State 1 -\textgreater{} State 3
\item
  Transition 3 - State 2 -\textgreater{} State 3
\end{itemize}

which is defined by the following matrix:

\begin{Shaded}
\begin{Highlighting}[]
\NormalTok{. }\FunctionTok{matrix}\NormalTok{ tmat = (.,1,2\textbackslash{}.,.,3\textbackslash{}.,.,.)}
\end{Highlighting}
\end{Shaded}

The key is to think of the column/row numbers as the states, and the
elements of the matrix as the transition numbers. Any transitions
indexed with a missing value \texttt{.} means that the transition
between the row state and the column state is not possible. Let's make
it obvious, sticking with our ``healthy'', ``ill'' and ``dead'' names
for the states:

\begin{Shaded}
\begin{Highlighting}[]
\NormalTok{. mat }\OtherTok{colnames}\NormalTok{ tmat = }\StringTok{"healthy"} \StringTok{"ill"} \StringTok{"dead"}

\NormalTok{. mat }\OtherTok{rownames}\NormalTok{ tmat = }\StringTok{"healthy"} \StringTok{"ill"} \StringTok{"dead"}

\NormalTok{. mat }\OtherTok{list}\NormalTok{ tmat}

\NormalTok{tmat[3,3]}
\NormalTok{         healthy      ill     dead}
\NormalTok{healthy        .        1        2}
\NormalTok{    ill        .        .        3}
\NormalTok{   dead        .        .        .}
\end{Highlighting}
\end{Shaded}

Now we've defined the transition matrix, we can use \texttt{survsim} to
simulate some data. We'll simulate 1000 observations, and generate a
binary treatment group indicator, remembering to \texttt{set\ seed}
first.

\begin{Shaded}
\begin{Highlighting}[]
\NormalTok{. }\KeywordTok{clear}

\NormalTok{. }\KeywordTok{set} \KeywordTok{obs}\NormalTok{ 1000}
\NormalTok{Number }\KeywordTok{of}\NormalTok{ observations (_N) was 0, now 1,000.}

\NormalTok{. }\KeywordTok{set} \DecValTok{seed}\NormalTok{ 9865}

\NormalTok{. }\KeywordTok{gen}\NormalTok{ trt = runiform()>0.5}
\end{Highlighting}
\end{Shaded}

The first transition-specific hazard has a user defined baseline hazard
function, with a harmful treatment effect. The second
transition-specific hazard model has a Weibull distribution, with a
beneficial treatment effect. The third transition-specific hazard has a
user-defined baseline hazard function, with an initially beneficial
treatment effect that reduces linearly with respect to log time. Right
censoring is applied at 3 years.

\begin{Shaded}
\begin{Highlighting}[]
\NormalTok{. survsim time state event, transmatrix(tmat)                  }\CommentTok{///}
\NormalTok{>         hazard1(user(}\FunctionTok{exp}\NormalTok{(-2 :+ 0.2:* }\FunctionTok{log}\NormalTok{(\{t\}) :+ 0.1:*\{t\}))  }\CommentTok{///}
\NormalTok{>                 covariates(trt 0.1))                         }\CommentTok{///}
\NormalTok{>         hazard2(dist(}\KeywordTok{weibull}\NormalTok{) lambda(0.01) }\KeywordTok{gamma}\NormalTok{(1.3)        }\CommentTok{///}
\NormalTok{>                 covariates(trt -0.5))                        }\CommentTok{///}
\NormalTok{>         hazard3(user(0.1 :* \{t\} :^ 1.5) covariates(trt -0.5) }\CommentTok{///}
\NormalTok{>                 tde(trt 0.1) tdefunction(}\FunctionTok{log}\NormalTok{(\{t\})))          }\CommentTok{///}
\NormalTok{>         maxtime(3)}
\NormalTok{variables time0 to time2 created}
\NormalTok{variables state0 to state2 created}
\NormalTok{variables event1 to event2 created}
\end{Highlighting}
\end{Shaded}

The hazard number \texttt{\#} in each \texttt{hazard\#()}, represents
the transition number in the transition matrix. Simple as that.
\texttt{survsim} has created variables storing the times at which states
were entered, with the associated state number and the associated event
indicator. It begins by creating the \texttt{0} variables, which
represents the time at which observations entered the initial state,
\texttt{time0}, and the associated state number, \texttt{state0}. As
\texttt{ltruncated()} and \texttt{startstate()} were not specified, all
observations are assumed to start in state 1 at time 0. Subsequent
transitions are simulated until all observations have either entered an
absorbing state, or are right-censored at their \texttt{maxtime()}. For
simplicity, I will assume time is measured in years. We can see what
\texttt{survsim} has created:

\begin{Shaded}
\begin{Highlighting}[]
\NormalTok{. }\OtherTok{list} \KeywordTok{if} \FunctionTok{inlist}\NormalTok{(}\DataTypeTok{_n}\NormalTok{,1,4,16,112)}

\NormalTok{      +-------------------------------------------------------------------------+}
\NormalTok{   1. | trt | time0 | state0 |     time1 | state1 | event1 |     time2 | state2 |}
\NormalTok{      |   0 |     0 |      1 |         3 |      1 |      0 |         . |      . |}
\NormalTok{      |-------------------------------------------------------------------------|}
\NormalTok{      |                                 event2                                  |}
\NormalTok{      |                                      .                                  |}
\NormalTok{      +-------------------------------------------------------------------------+}

\NormalTok{      +-------------------------------------------------------------------------+}
\NormalTok{   4. | trt | time0 | state0 |     time1 | state1 | event1 |     time2 | state2 |}
\NormalTok{      |   1 |     0 |      1 | .95636156 |      2 |      1 |         3 |      2 |}
\NormalTok{      |-------------------------------------------------------------------------|}
\NormalTok{      |                                 event2                                  |}
\NormalTok{      |                                      0                                  |}
\NormalTok{      +-------------------------------------------------------------------------+}

\NormalTok{      +-------------------------------------------------------------------------+}
\NormalTok{  16. | trt | time0 | state0 |     time1 | state1 | event1 |     time2 | state2 |}
\NormalTok{      |   0 |     0 |      1 | 1.0755764 |      2 |      1 | 2.4401409 |      3 |}
\NormalTok{      |-------------------------------------------------------------------------|}
\NormalTok{      |                                 event2                                  |}
\NormalTok{      |                                      1                                  |}
\NormalTok{      +-------------------------------------------------------------------------+}

\NormalTok{      +-------------------------------------------------------------------------+}
\NormalTok{ 112. | trt | time0 | state0 |     time1 | state1 | event1 |     time2 | state2 |}
\NormalTok{      |   1 |     0 |      1 | 2.3290322 |      3 |      1 |         . |      . |}
\NormalTok{      |-------------------------------------------------------------------------|}
\NormalTok{      |                                 event2                                  |}
\NormalTok{      |                                      .                                  |}
\NormalTok{      +-------------------------------------------------------------------------+}
\end{Highlighting}
\end{Shaded}

All observations start initially in state 1 at time 0, which are stored
in \texttt{state0} and \texttt{time0}, respectively. Then,

\begin{itemize}
\tightlist
\item
  Observation 1 is right-censored at 3 years, remaining in state 1
\item
  Observation 4 moves to state 2 at 0.956 years, and is subsequently
  right-censored at 3 years, still in state 2
\item
  Observation 16 moves to state 2 at 1.076 years, and then moves to
  state 3 at 2.440 years. Since state 3 is an absorbing state, there are
  no further transitions
\item
  Observation 112 moves to state 3 at 2.329 years. Again, since state 3
  is absorbing, there are no further transitions
\end{itemize}

There's a variety of extensions we could incorporate, for example, we
could simulate from a semi-Markov model by using the \texttt{reset}
option in \texttt{hazard3()}, which would reset the clock when State 2
is entered. The simulated event times that \texttt{survsim} returns will
still be calculated on the main timescale in this case, time since
initial \texttt{startstate()}. We could of course have a much more
complex multi-state structure, i.e.~more states or reversible
transitions. Both of these are supported by \texttt{survsim}.

\subsection{Simulating from an illness-death model with multiple timescales}

A further capability of \texttt{survsim} is to simulate from an event
time model with \emph{multiple} timescales. Such a setting is rarely
considered, as both the estimation of such a model is computationally
challenging, let alone simulating from one. However, \texttt{survsim}
now provides this, in extremely general way, and indeed \texttt{merlin}
has the capability of fitting such models.

Continuing with the illness-death model, the transition between the
illness and death state may depend not only on time since the initial
healthy state entry, but also on the time \emph{since} illness. More
formally, we can define the transition rate for state 2 to 3 as,

\[
h_{3}(t) = h_{30}(t) \exp (X\beta_{31} + f(t-t_{30})\beta_{32})
\]

where \(h_{30}(t)\) is the baseline hazard function on the main
timescale, time since entry to the healthy state 1. We have a vector of
baseline covariates with associated log hazard ratios, \(X\) and
\(\beta_{31}\), respectively. Finally, we define \(t_{30}\) to be the
time at which the observation entered State 2, and hence \((t-t_{30})\)
defines the time since entry to state 2. Our function \(f()\) can be as
simple or complex as required, with associated coefficient vector
\(\beta_{32}\).

For simplicity, I'll assume a linear effect of time since entry to state
2, incorporating it into the illness-death transition models from the
previous section, setting \(\beta_{32} = -0.05\), meaning a longer time
to illness reduces the risk of death.

\begin{Shaded}
\begin{Highlighting}[]
\NormalTok{. }\KeywordTok{capture} \KeywordTok{drop}\NormalTok{ time* state* event*}

\NormalTok{. }\KeywordTok{set} \DecValTok{seed}\NormalTok{ 98798}

\NormalTok{. survsim time state event, transmatrix(tmat)                           }\CommentTok{///}
\NormalTok{>         hazard1(user(}\FunctionTok{exp}\NormalTok{(-2 :+ 0.2:* }\FunctionTok{log}\NormalTok{(\{t\}) :+ 0.1:*\{t\}))           }\CommentTok{///}
\NormalTok{>                 covariates(trt 0.1))                                  }\CommentTok{///}
\NormalTok{>         hazard2(dist(}\KeywordTok{weibull}\NormalTok{) lambda(0.01) }\KeywordTok{gamma}\NormalTok{(1.3)                 }\CommentTok{///}
\NormalTok{>                 covariates(trt -0.5))                                 }\CommentTok{///}
\NormalTok{>         hazard3(user(0.1 :* \{t\} :^ 1.5 :* }\FunctionTok{exp}\NormalTok{(-0.05 :* (\{t\}:-\{t0\})))  }\CommentTok{///}
\NormalTok{>                 covariates(trt -0.5)                                  }\CommentTok{///}
\NormalTok{>                 tde(trt 0.1) tdefunction(}\FunctionTok{log}\NormalTok{(\{t\})))                   }\CommentTok{///}
\NormalTok{>         maxtime(3)}
\NormalTok{variables time0 to time2 created}
\NormalTok{variables state0 to state2 created}
\NormalTok{variables event1 to event2 created}
\end{Highlighting}
\end{Shaded}

To refer to the entry time for a particular transition, we use
\texttt{\{t0\}} alongside our usual \texttt{\{t\}} denoting time since
study origin, and hence \texttt{(\{t\}\ :-\ \{t0\})} allows us to define
time since state entry. Of course, this can be extended in numerous
ways.

\subsection{Simulating from a reversible illness-death model}

I now extend the example from Section \ref{sec:illdeg} to allow for
recovery, i.e.~observations can move from the ``illness'' state back to
the ``healthy state''. Our transition matrix now looks like this:

\begin{Shaded}
\begin{Highlighting}[]
\NormalTok{. }\FunctionTok{matrix}\NormalTok{ tmat = (.,1,2\textbackslash{}3,.,4\textbackslash{}.,.,.)}

\NormalTok{. mat }\OtherTok{colnames}\NormalTok{ tmat = }\StringTok{"healthy"} \StringTok{"ill"} \StringTok{"dead"}

\NormalTok{. mat }\OtherTok{rownames}\NormalTok{ tmat = }\StringTok{"healthy"} \StringTok{"ill"} \StringTok{"dead"}

\NormalTok{. }\FunctionTok{matrix} \OtherTok{list}\NormalTok{ tmat}

\NormalTok{tmat[3,3]}
\NormalTok{         healthy      ill     dead}
\NormalTok{healthy        .        1        2}
\NormalTok{    ill        3        .        4}
\NormalTok{   dead        .        .        .}
\end{Highlighting}
\end{Shaded}

Remember, transitions must be indexed going from top left to bottom
right, along columns then down rows. So transition 3 is our new
transition, going from ``ill'' to ``healthy''.

Once again we'll simulate 1000 observations, and generate a binary
treatment group indicator, remembering to \texttt{set\ seed} first.
However, now I''ll add delayed entry, allowing each observation to enter
the study at a time greater than 0. Remember, entry and transition times
are all simulated and recorded on the main overall timescale, even if
the clock is \texttt{reset} on state entry. I will also vary the
starting state of the observations, allowing observations to enter in
either state 1 or 2.

\begin{Shaded}
\begin{Highlighting}[]
\NormalTok{. }\KeywordTok{clear}

\NormalTok{. }\KeywordTok{set} \KeywordTok{obs}\NormalTok{ 1000}
\NormalTok{Number }\KeywordTok{of}\NormalTok{ observations (_N) was 0, now 1,000.}

\NormalTok{. }\KeywordTok{set} \DecValTok{seed}\NormalTok{ 9865}

\NormalTok{. }\KeywordTok{gen}\NormalTok{ trt = runiform()>0.5}

\NormalTok{. }\KeywordTok{gen}\NormalTok{ lt = 1.5 * runiform()}

\NormalTok{. }\KeywordTok{gen}\NormalTok{ startstate = 1 + (runiform()>0.5)}
\end{Highlighting}
\end{Shaded}

The first transition-specific hazard has a user defined baseline hazard
function, with a harmful treatment effect. The second
transition-specific hazard model has a Weibull distribution, with a
beneficial treatment effect. The third transition has a Weibull
distribution with no covariate effects. The fourth transition-specific
hazard has a user-defined baseline hazard function, with an initially
beneficial treatment effect that reduces linearly with respect to log
time. Right censoring is applied at 3 years.

\begin{Shaded}
\begin{Highlighting}[]
\NormalTok{. survsim time state event, transmatrix(tmat)                  }\CommentTok{///}
\NormalTok{>         hazard1(user(}\FunctionTok{exp}\NormalTok{(-2 :+ 0.2:* }\FunctionTok{log}\NormalTok{(\{t\}) :+ 0.1:*\{t\}))  }\CommentTok{///}
\NormalTok{>                 covariates(trt 0.1))                         }\CommentTok{///}
\NormalTok{>         hazard2(dist(}\KeywordTok{weibull}\NormalTok{) lambda(0.01) }\KeywordTok{gamma}\NormalTok{(1.3)        }\CommentTok{///}
\NormalTok{>                 covariates(trt -0.5))                        }\CommentTok{///}
\NormalTok{>         hazard3(dist(}\KeywordTok{weibull}\NormalTok{) lambda(0.05) }\KeywordTok{gamma}\NormalTok{(1))         }\CommentTok{///}
\NormalTok{>         hazard4(user(0.1 :* \{t\} :^ 1.5) covariates(trt -0.5) }\CommentTok{///}
\NormalTok{>                 tde(trt 0.1) tdefunction(}\FunctionTok{log}\NormalTok{(\{t\})))          }\CommentTok{///}
\NormalTok{>         ltruncated(lt) startstate(startstate) maxtime(3)}
\NormalTok{variables time0 to time4 created}
\NormalTok{variables state0 to state4 created}
\NormalTok{variables event1 to event4 created}
\end{Highlighting}
\end{Shaded}

As \texttt{ltruncated()} and \texttt{startstate()} were specified,
observations start in different states and at different times.
Subsequent transitions are simulated until all observations have either
entered an absorbing state, or are right-censored at their
\texttt{maxtime()}. For simplicity, I will assume time is measured in
years. We can see what \texttt{survsim} has created:

\begin{Shaded}
\begin{Highlighting}[]
\NormalTok{. }\OtherTok{list} \KeywordTok{if} \FunctionTok{inlist}\NormalTok{(}\DataTypeTok{_n}\NormalTok{,1,695,704)}

\NormalTok{      +---------------------------------------------------------------------+}
\NormalTok{   1. | trt |       lt | starts~}\FunctionTok{e}\NormalTok{ |     time0 | state0 |     time1 | state1 |}
\NormalTok{      |   0 | .0026511 |        1 | .00265108 |      1 |         3 |      1 |}
\NormalTok{      |---------------------------------------------------------------------|}
\NormalTok{      | event1 |     time2 | state2 | event2 |     time3 | state3 | event3  |}
\NormalTok{      |      0 |         . |      . |      . |         . |      . |      .  |}
\NormalTok{      |---------------------------------------------------------------------|}
\NormalTok{      |        time4        |        state4         |        event4         |}
\NormalTok{      |            .        |             .         |             .         |}
\NormalTok{      +---------------------------------------------------------------------+}

\NormalTok{      +---------------------------------------------------------------------+}
\NormalTok{ 695. | trt |       lt | starts~}\FunctionTok{e}\NormalTok{ |     time0 | state0 |     time1 | state1 |}
\NormalTok{      |   1 | .7018383 |        2 | .70183825 |      2 |  1.244147 |      1 |}
\NormalTok{      |---------------------------------------------------------------------|}
\NormalTok{      | event1 |     time2 | state2 | event2 |     time3 | state3 | event3  |}
\NormalTok{      |      1 | 1.9200901 |      2 |      1 | 2.2587202 |      1 |      1  |}
\NormalTok{      |---------------------------------------------------------------------|}
\NormalTok{      |        time4        |        state4         |        event4         |}
\NormalTok{      |            3        |             1         |             0         |}
\NormalTok{      +---------------------------------------------------------------------+}

\NormalTok{      +---------------------------------------------------------------------+}
\NormalTok{ 704. | trt |       lt | starts~}\FunctionTok{e}\NormalTok{ |     time0 | state0 |     time1 | state1 |}
\NormalTok{      |   0 | 1.279466 |        2 | 1.2794662 |      2 | 2.3864832 |      1 |}
\NormalTok{      |---------------------------------------------------------------------|}
\NormalTok{      | event1 |     time2 | state2 | event2 |     time3 | state3 | event3  |}
\NormalTok{      |      1 |  2.464492 |      2 |      1 | 2.9616442 |      3 |      1  |}
\NormalTok{      |---------------------------------------------------------------------|}
\NormalTok{      |        time4        |        state4         |        event4         |}
\NormalTok{      |            .        |             .         |             .         |}
\NormalTok{      +---------------------------------------------------------------------+}
\end{Highlighting}
\end{Shaded}

Starting state and times are stored in \texttt{state0} and
\texttt{time0}, respectively. We see that,

\begin{itemize}
\tightlist
\item
  Observation 1 enters state 1 at 0.00265 years and remains there until
  they are right-censored at 3 years
\item
  Observation 695 enters state 2 at 0.7018 years, recovers to state 1 at
  1.244 years, has a relapse at 1.920 years moving to state 2, recovers
  again at 2.259 years, and remains there until the end of follow-up.
\item
  Observation 704 enters state 2 at 1.279 years, recovers to state 1 at
  2.386 years, has a relapse at 2.464 years moving to state 2, and
  subsequently dies at 2.9616 years, moving to state 3.
\end{itemize}

Note in this case we had 4 sets of new variables created. If we extend
follow-up, then we will obtain more since we have a reversible process,
and \texttt{survsim} will simply continue to simulate transitions for
observations until all observations have reached an absorbing state, or
the maximum follow-up time.

\section{Conclusion}
\label{sec:conc}

In this paper I have introduced a range of extensions to the
\texttt{survsim} command for simulating survival data from parametric
distributions, custom/user-defined hazard or cumulative hazard
functions, an estimated \texttt{merlin} survival model, or from a
general competing risks or multi-state model.

For those interested in competing risks and multi-state modelling, I
refer you to the \texttt{multistate} package, where for example, the
simulated datasets shown above may be \texttt{msset} and subsequently
analysed within a multi-state setting \citep{Crowther2017f}.

Further work will allow the ability to simulate from multivariate and
hierarchical/multilevel \texttt{merlin} models.

\section*{About the author}

Michael J. Crowther is a Biostatistician at Karolinska Institutet and a
Consultant Stata Developer. He works heavily in methods and software
development, particularly in the field of survival analysis.

\renewcommand\refname{References}
  \bibliography{survsim_paper2.bbl}

\end{document}